\documentclass[aps,prd,preprint,tightenlines,groupedaddress,showpacs]{revtex4}
\usepackage{epsf,epsfig,graphics,graphicx}
\def\edth{\;\raise1.0pt\hbox{$'$}\hskip-6pt\partial\;}
\def\baredth{\;\overline{\raise1.0pt\hbox{$'$}\hskip-6pt
\partial}\;}
\def\gsim{~\rlap{$>$}{\lower 1.0ex\hbox{$\sim$}}}

\newcommand{\be}{\begin{equation}}
\newcommand{\ba}{\begin{eqnarray}}
\newcommand{\ee}{\end{equation}}
\newcommand{\ea}{\end{eqnarray}}

\newcommand{\fr}{\frac}
\newcommand{\tF}{\tilde{F}}

\begin{document}

\title{Dark Ultra-Light Scalars and Cosmic Parity Violation}

\author{Seokcheon Lee$^{1}$, Guo-Chin Liu$^{2}$, and Kin-Wang Ng$^{3,4}$}

\affiliation{
$^1$Research Institute of Natural Science, Gyeongsang National University, Jinju 52828, Korea\\
$^2$Department of Physics, Tamkang University, Tamsui, New Taipei City 25137, Taiwan\\
$^3$Institute of Physics, Academia Sinica, Taipei 11529, Taiwan\\
$^4$Institute of Astronomy and Astrophysics, Academia Sinica, Taipei 11529, Taiwan
}

\begin{abstract}
If the dark sector of the Universe consists of ultra-light scalars, their coupling to photon via a Chern-Simons term
would induce a rotation of the polarization plane of the cosmic microwave background (CMB).
This rotation would convert $E$-mode polarization into $B$-mode polarization, resulting in new
CMB $BB$ correlation and parity-violating $TB$ and $EB$ cross correlations. We review the subject giving details about
the derivation of the rotational effects and summarizing the possible signals in current and future CMB $B$-mode experiments.
\end{abstract}

\pacs{95.35.+d, 95.36.+x, 98.70.Vc}
\maketitle

\section{Introduction}

The existence of a dark sector is concordantly supported by many astrophysical
and cosmological observations~\cite{planck16}. The dark sector has been successfully treated
as a combination of pressure-less dark matter (DM) and dark energy (DE) with negative pressure.
A lot of effort has been put to measure the properties of the dark components. Nevertheless,
they remain elusive and are indeed the most mysterious matter that we have ever imagined.
To understand the microscopic nature as well as to measure
cosmological signals of the dark components have become among
the most important goals in cosmological research.

Cosmological constant is the simplest explanation for the existence of DE. 
Unfortunately, the observed value of the cosmological
constant is totally mismatched with the theoretical
expectation~\cite{Weinberg}. An alternative candidate, described
by a dynamical scalar field $\Phi$, is thus considered. The
dynamics of $\Phi$ is governed by a scalar potential $V(\Phi)$ with a canonical kinetic
term (called quintessence) or a modified one allowing negative
kinetic energy (called phantom), which whatsoever makes the DE dominant in the recent epoch. 
Last but not least, other scalar DE models than quintessence and phantom have been
proposed~\cite{DErev}. In this review, we will consider only the
quintessence models. Indeed, our method can be easily applied to
other scalar models. There are many different kinds of 
quintessential potentials, for example, the pseudo Nambu-Goldstone
boson, inverse power law, exponential, hyperbolic cosine, and
tracking oscillating~\cite{potential}. To differentiate between
these models and finally reconstruct $V(\Phi)$ would likely require
next-generation observations. 

The quintessential potential $V(\Phi)$ and the field $\Phi$ itself
are difficult to be measured directly.  What we can do is to
investigate the DE energy density $\rho_\Phi$
and the time evolution of the DE pressure $p_\Phi$ or
the equation of state (EOS) $w_\Phi\equiv p_\Phi/\rho_\Phi$,
both of which are governed by the dynamics of $\Phi$.
Several observations, such as the 157 supernovae in a redshift interval, $0.015 < z < 1.6$, 
in the ``Gold Sample'' obtained from a combination of ground-based data
and the Hubble Space Telescope~\cite{Riess} and the 115 supernovae
with $0.015< z < 1$ from the Supernova Legacy Survey, have provided
constraints on the DE EOS~\cite{snls}.  Joint analysis of CMB data
with supernovae or/and large scale structure survey such as SDSS
or 2dfGRS can offer better constraints on quintessence
models~\cite{joint}. Furthermore, the study of the cross-correlation
of maps between CMB and various tracers of matter through the
integrated Sachs-Wolfe effect has also been carried out~\cite{isw}. 
However, the time evolution of DE is still poorly constrained by observations, 
thus allowing a very wild range of the DE EOS, which is strongly model dependent.

Although the nature of DM remains unknown,
its gravitational pull is essential to the formation of large-scale structures.
It has been successfully modeled as massive weakly interacting particles or cold dark matter (CDM).
However, there exist serious discrepancies between observations and numerical simulations of CDM halos,
which predict too much power on small scales, manifested as cuspy CDM cores in dwarf galaxies,
galaxies like the Milky Way, and central regions of galaxy clusters
as well as a large excess of CDM subhalos or dwarf galaxies.
These discrepancies, if true, would suggest a suppressed matter power spectrum at small scales~\cite{primack}.
Massive scalar particle is a viable candidate for CDM. As long as the condition $m > 3H$, where $m$ is the scalar mass and $H$ is the Hubble parameter, is satisfied, the scalar begins to coherently oscillate with an amplitude set by its initial vacuum expectation value (vev). This constitutes a homogeneous condensate with its energy density redshifting as $a^{-3}$ (where $a$ is the cosmic scale factor). If $m > 10^{-27}{\rm eV}$, the scalar condensate behaves just like CDM after matter-radiation equality. 
Interestingly, for ultra-light scalars with masses $m < 10^{-20}{\rm eV}$, the de Broglie wave can suppress small-scale power on astronomically observable length scales~\cite{hu,hlozek,huilam}. In numerical calculations of the scalar field with $m \sim 10^{-22}{\rm eV}$, it was shown that the scalar model may offer a viable solution to the small-scale problems~\cite{hu,marsh}. In this review, we will look at this ultra-light scalar DM.

Although the physical state of the dark sector can be measured through
their gravitational effects on the evolution of the cosmological
background, it is important to know their microscopic nature.
It will be a breakthrough in fundamental physics if we
find that DE is not a cosmological constant,
but rather a nearly massless, slowly rolling scalar field, or that DM is a scalar condensate.
Interestingly, they may be the other fundamental scalars that already exist in nature 
after the discovery of the Higgs boson in the Standard Model at
the Large Hadron Collider~\cite{LHC}.
A possible way to probe the nature of scalar DM and DE is to study the interaction of
the scalar $\Phi$ to ordinary matter. Recently, there have been a lot of interests in
studying the observational effects of direct interaction of $\Phi$
to electromagnetism, which include the rotation of polarized light
from distant radio sources~\cite{Carroll1998} and the generation
of large-scale magnetic fields for a pseudo-scalar-type coupling~\cite{Lee02},
as well as the temporal evolution of the fine structure constant for a scalar-type
coupling~\cite{fine}. Here we will concentrate on the pseudo-scalar-type coupling or
the Chern-Simons term, which leads to the rotation of the polarization plane of the CMB,
converting $E$-mode into $B$-mode polarization without affecting the temperature anisotropy ($T$)~\cite{Lue,LLN,LLN14,LN17}.
This results in new CMB $BB$ correlation and parity-violating $TB$ and $EB$ cross correlations, which
can be tested in CMB $B$-mode experiments~\cite{planckCPV}.

The review is organized as follows. In the next section, we will
introduce the pseudo-scalar $\Phi$-photon coupling. The radiative
transfer equation for CMB in dark scalar cosmology is derived
in Sec.~\ref{QUsec}, and its analytic approximated solution is presented in Sec.~\ref{freestream}.
Section~\ref{powerspec} shows the power
spectra for the CMB temperature and polarization anisotropies. In
this section, we will also discuss the implication of the results
to on-going and future experiments in search of CMB primordial and lensing $B$-mode polarization. 
Section~\ref{conclus} is our conclusion.

\section{Pseudo-scalar Coupling of Dark Sector to Photon}

Three decades ago Ni found the most general interaction Lagrangian
for an electromagnetic system in a gravitational field which was
introduced as a unique counterexample to Schiff's conjecture (
{\it i.e.}, any consistent Lorentz invariant theory of gravity
which obeys the weak equivalence principle would necessarily obey
the Einstein equivalence principle or the minimal coupling metric
theory of gravity)~\cite{Ni}:
\be
{\cal L}_{N} = - \fr{1}{4}
\sqrt{-g} B_{F \tilde{F}}(\phi) F_{\mu\nu} \tF^{\mu \nu}\, ,
\quad {\rm where}\quad\phi\equiv \fr{\Phi}{M} \label{LN} \, ,
\ee
where $g$ is the determinant of the metric, the
electromagnetic field strength tensor is
$F_{\mu \nu} = \nabla_{\mu} A_{\nu} - \nabla_{\nu} A_{\mu} =
\partial_{\mu} A_{\nu} - \partial_{\nu} A_{\mu}$, and its dual is
given by $\tilde{F}^{\mu \nu} = \fr{1}{2} \epsilon^{\mu \nu \rho
\sigma} F_{\rho \sigma}/\sqrt{-g}$.
Note that $1/\sqrt{-g}$ is added to the dual tensor because
$\epsilon^{\mu \nu \rho \sigma}$ is a tensor density of weight
$-1$~\cite{wein}.
Here we have introduced the reduced Planck mass $M=M_{Pl}/\sqrt{8\pi}$, scaled
$\Phi$ to a dimensionless field $\phi$, and assumed the scalar
function in Ref.~\cite{Ni} as a $\phi$-dependent dimensionless
coupling term $B_{F \tilde{F}}(\phi)$. He also suggested that this
non-metric theory of gravity may be related to the existence of
parity-non-conserving field or spontaneously broken symmetry. A
concrete example that pseudoscalar or axion-like couplings arise from the
spontaneous breaking of a compact symmetry group, say, U(1) can be
found in Frieman {\it et al.} in Ref.~\cite{potential}. Recently, it was proposed that string theory
suggests the presence of a plenitude of axions (an axiverse),
possibly populating each decade of mass down to the
Hubble scale, and naturally coupled to photons~\cite{arvanitaki}. Models
of string axions as candidates for dark energy have also been discussed~\cite{panda}.
Thus, the mean field or the vev of an axion-like $\phi$ explicitly breaks the parity symmetry; instead, $\phi$ can be
treated as a scalar that has a parity-violating $\phi$-photon interaction.
In scalar DM or DE models, the time-evolving
mean field $\langle\phi\rangle \equiv {\bar\phi}$ generates rotation-induced CMB $B$-mode polarization
with parity-violating $TB$ and $EB$ correlations. In addition, the field perturbation $\delta\phi$ induces
a new $BB$ correlation while preserving parity. 

We can write the general action of a scalar field in Einstein gravity
including the electromagnetic interaction and Ni interaction as
\be
S = \int d^4
x \sqrt{-g} \Biggl[ \fr{M^2}{2} \Bigl( R - \nabla^{\mu} \phi
\nabla_{\mu} \phi \Bigr) - V(\phi) - \fr{1}{4} F_{\mu\nu}
F^{\mu\nu} + \fr{{\cal L}_{N}}{\sqrt{-g}} + \cdots \Biggr]
\label{S} \, ,
\ee
where $V(\phi)$ is the self-interaction potential of the
scalar field and $1/4 \sqrt{-g} F_{\mu\nu} F^{\mu\nu}$ is the
free Maxwell Lagrangian density. We can integrate by parts the Ni
Lagrangian density to get
\ba
{\cal L}_{N} &=& - \fr{1}{4} \sqrt{-g} B_{F \tilde{F}} F_{\mu \nu} \tilde{F}^{\mu \nu}
= -\fr{1}{2}\sqrt{-g} B_{F \tilde{F}} (\nabla_{\mu} A_{\nu}) \tF^{\mu\nu} \nonumber \\
&=& \fr{1}{2}\sqrt{-g} \Bigl[ - \nabla_{\mu} (B_{F \tilde{F}} A_{\nu}
\tF^{\mu\nu}) + \nabla_{\mu} (B_{F \tilde{F}} \tF^{\mu\nu})
A_{\nu} \Bigr] = \fr{1}{2}\sqrt{-g} \Bigl[ \nabla_{\mu} (B_{F \tilde{F}}
\tF^{\mu\nu}) A_{\nu} \Bigr] \label{LN2} \, ,
\ea
where we have used the fact that $\tF^{\mu\nu}$ is antisymmetric
in the second equality and we have
ignored the divergence in the last equality. We can repeat the same
process for the Maxwell Lagrangian density and obtain
\ba
{\cal L}_{M} &=& -
\fr{1}{4} \sqrt{-g} F_{\mu\nu}F^{\mu\nu} = -
\fr{1}{2} \sqrt{-g} \nabla_{\mu}(A_{\nu}) F^{\mu\nu} \nonumber \\
&=& \fr{1}{2} \sqrt{-g} \Biggl[ -\nabla_{\mu} (A_{\nu}F^{\mu\nu})
+ \nabla_{\mu}(F^{\mu\nu}) A_{\nu} \Biggr] = \fr{1}{2} \sqrt{-g}
[\nabla_{\mu}(F^{\mu\nu}) A_{\nu}] \label{LM} \, .
\ea
From the above action (\ref{S}) we have the following equation which gives
the constraint equation for the interaction between the scalar
field and the electromagnetic field,
\be
M^2  \nabla_{\mu} \nabla^{\mu} \phi -
\fr{\partial V}{\partial \phi} - \fr{1}{4}\fr{\partial
B_{F\tilde{F}}}{\partial \phi} F_{\mu \nu} \tilde{F}^{\mu\nu} = 0
\label{phieq} \, .
\ee
We will consider the effect of the perturbation of
the scalar field. However, we will assume that the
back-reaction is negligible in the field equation, {\it i.e.},
$\partial V/ \partial \phi \gg (\partial B_{F \tF} / \partial
\phi) F \tF$. We can find the field equations for the
electromagnetic field by considering only Maxwell and Ni
Lagrangian densities from the action~(\ref{S}),
\ba
S_{M + N} &=& - \int d^4x \sqrt{-g} \Biggl[ \fr{1}{4} F_{\mu\nu} F^{\mu\nu} +
\fr{1}{4} B_{F\tF} F_{\mu\nu} \tF^{\mu\nu} \Biggr]  \nonumber \\
&=& \int d^4x \sqrt{-g} \Biggl[\fr{1}{2} \nabla_{\mu}(F^{\mu\nu}) A_{\nu} +\fr{1}{2}
\nabla_{\mu}(B_{F\tF} \tF^{\mu\nu}) A_{\nu} \Biggr] \label{SMN} \,,
\ea
where we have used Eqs.~(\ref{LN2}) and~(\ref{LM}). From this equation
with the Bianchi identity, we obtain the field equations as
\ba
\nabla_{\mu} F^{\mu\nu} &=& - \nabla_{\mu} (B_{F \tF})
\tF^{\mu\nu} \label{nF} \, , \\ \nabla_{\mu} \tF^{\mu\nu} &=& 0
\label{ntF} \, .
\ea
In the following, we will assume a flat Robertson-Walker metric:
\be
ds^2=-g_{\mu\nu} dx^\mu dx^\nu= a^2(\eta) (d\eta^2- d \vec{x}^2)
\label{metric} \, ,
\ee
where $a(\eta)$ is the cosmic scale factor and $\eta$ is the conformal time
defined by $dt=a(\eta)d\eta$.
It is well known that the minimal coupling of photons to
the metric background is conformally invariant. As
such, in the conformally flat metric~(\ref{metric}), it is convenient to
work with the conformal time for solving Eqs.~(\ref{nF}) and~(\ref{ntF}),
where we then have $\nabla_{\mu}=\partial_\mu=(\partial/\partial_\eta,\vec{\nabla})$.
Let us write them explicitly in terms of $\vec{E}$ and $\vec{B}$,
\ba
\vec{\nabla} \times \vec{B} - \fr{\partial \vec{E}}{\partial \eta} &=&
\fr{\partial B_{F\tF}}{\partial \phi} \fr{\partial \phi}{\partial
\eta} \vec{B} +  \fr{\partial B_{F\tF}}{\partial \phi} \vec{\nabla}
\phi \times \vec{E} \label{nFi} \, , \\
\vec{\nabla} \cdot \vec{E} &=& -  \fr{\partial B_{F\tF}}
{\partial \phi} \vec{\nabla} \phi \cdot \vec{B} \label{nF0} \, ,\\
\fr{\partial \vec{B}}{\partial \eta} + \vec{\nabla} \times \vec{E} &=& 0 \label{ntFi}\, , \\
\vec{\nabla} \cdot \vec{B} &=& 0 \label{ntF0} \, .
\ea
If we choose the temporal gauge for the four vector potential, namely
$A_{\mu} = (0, \vec{A})$, then we can rewrite Eqs.~(\ref{nFi})
and~(\ref{nF0}) as
\ba
\fr{\partial^2 \vec{A}}{\partial \eta^2} - \vec{\nabla}^2 \vec{A}
+ \vec{\nabla} (\vec{\nabla}\cdot \vec{A}) &=&  \fr{\partial B_{F\tF}}{\partial \phi}
\fr{\partial \phi}{\partial \eta} \vec{\nabla} \times \vec{A} -
\fr{\partial B_{F\tF}}{\partial \phi} \vec{\nabla} \phi \times
\fr{\partial \vec{A}}{\partial \eta} \label{nFi2} \, , \\
\fr{\partial}{\partial \eta}(\vec{\nabla} \cdot \vec{A}) &=&  \fr{\partial B_{F\tF}}
{\partial \phi} \vec{\nabla} \phi \cdot(\vec{\nabla} \times \vec{A}) \label{nF02} \, ,
\ea
where we have used $\vec{E} = - \partial \vec{A} / \partial \eta$
and $\vec{B} = \vec{\nabla} \times \vec{A}$.
To find wave solutions to these equations, we posit the ansatz,
\be
\vec{A}\propto e^{-in^\mu x_\mu}\propto e^{-i\varepsilon\eta+i\vec{n}\cdot\vec{x}}\, ,
\ee
where $n^\mu=(\varepsilon,\vec{n})$ is the photon four-momentum.
Hence we obtain a single wave equation in the Fourier space as
\be
\varepsilon^2\vec{A} - n^2 \vec{A}
+ \vec{n} (\vec{n}\cdot \vec{A}) = -i (p_0 \vec{n} \times \vec{A}
+ \varepsilon \vec{p} \times\vec{A})\, ,\quad{\rm where}\quad
p_\mu=(p_0,\vec p)\equiv \fr{\partial B_{F\tF}}{\partial \phi} \partial_\mu \phi\, .
\ee
By choosing $\vec{n}=n \hat{n}$ with $\hat{n}\parallel \hat{e}_3$ and decomposing
$\vec{A}=A_1 \hat{e}_1+A_2 \hat{e}_2 + A_3 \hat{e}_3$ and
$\vec{p}=p_1 \hat{e}_1+p_2 \hat{e}_2 + p_3 \hat{e}_3$, we get
$A_3=i(p_2 A_1-p_1 A_2)/\varepsilon$ and
\be
\left(
\begin{array}{cc}
  \varepsilon^2-n^2-p_2^2 &  -i(np_0+\varepsilon p_3+ip_1 p_2) \\
  i(np_0+\varepsilon p_3-ip_1 p_2) & \varepsilon^2-n^2-p_1^2 \\
\end{array}
\right)
\left(
\begin{array}{c}
  A_1\\
  A_2 \\
\end{array}
\right)
=0\, .
\ee
The determinant of this equation gives the dispersion relation,
\be
(\varepsilon^2-n^2)^2-(\varepsilon^2-n^2)p^2\sin^2\theta=(np_0+\varepsilon p\cos\theta)^2
\label{dispersion}\, ,
\ee
where $n=|\vec{n}|$, $p=|\vec{p}|$, and $\theta$
is the angle between $\vec{n}$ and $\vec{p}$, or in a covariant form,
\be
(n^\mu n_\mu)^2+(n^\mu n_\mu)(p^\mu p_\mu)=(n^\mu p_\mu)^2\, .
\ee
This dispersion relation with $p_\mu$ being considered as a constant
external four-vector was first derived in the model involving
a Lorentz- and parity-violating modification of electromagnetism~\cite{carroll90}.
Here we are considering a dynamical scalar field.
For photon with frequency $\varepsilon$ propagating in the $\hat{n}$ direction
in the presence of non-zero field four-gradient
$(\partial B_{F\tF}/\partial \phi)\partial_\mu \phi$, Eq.~(\ref{dispersion})
has two positive roots $n_\pm$ which correspond respectively to the wave numbers
of the left- and right-handed circularly polarized waves, $A_1\pm iA_2$.
It will prove to be useful to rewrite Eq.~(\ref{dispersion}) as
\be
\varepsilon^2-n^2=\pm(np_0+\varepsilon p\cos\theta)
\left[1-\frac{p^2\sin^2\theta}{\varepsilon^2-n^2}\right]^{-\frac{1}{2}}\, .
\ee
Since we expect that the coupling of the dark sector to photon is small,
we can expand this equation in powers of $p_0$ and $p$
to obtain, to first order,
\be
n_\pm=\varepsilon\mp{1\over2}\fr{\partial B_{F\tF}}{\partial \phi}
\left(\fr{\partial \phi}{\partial\eta}+
\vec{\nabla} \phi \cdot \hat{n}\right).
\ee
As a consequence, the dispersion leads to a rotation of
the polarization plane with angular velocity equal to
\be
\omega=\frac{1}{2}(n_+ - n_-)=
-{1\over2}\fr{\partial B_{F\tF}}{\partial \phi}
\left(\fr{\partial \phi}{\partial\eta}+
\vec{\nabla} \phi \cdot \hat{n}\right).
\label{omegaetax}
\ee
Hence the rotational velocity is $\omega(\eta, \vec{x})$ that depends
both on the time variation of the field $\phi(\eta,\vec{x})$ and its
spatial gradient in the propagating direction of the photon.
Assuming a spatially homogeneous field $\phi(\eta,\vec{x}) = \bar{\phi}(\eta)$
and $B_{F\tF} = \beta_{F\tF} \phi$ with $\beta_{F\tF}$ being a constant, we have
\be
{\bar\omega}(\eta)= -{1\over2} \beta_{F\tF} \fr{d\bar{\phi}}{d\eta}.
\label{omegaeta}
\ee

Carroll~\cite{Carroll1998} proposed that Eq.~(\ref{omegaeta}) would lead to a rotation
of the polarization direction of light from distant radio galaxies and quasars
if the spatially homogeneous quintessence slowly varies with time.
This effect is called as the ``cosmological birefringence''.
The rotated angle of the polarization direction for an observed
source would then be given by
\be
{\bar\alpha}=\int_z^0 {\bar\omega}(\eta) d\eta=
-{1\over2} \beta_{F\tF} \Delta \bar{\phi},
\ee
where $\Delta \bar{\phi}$ is the change in $\bar{\phi}$ between
the redshift $z$ of the source and today. Measurements of the rotated angles
for distant astrophysical sources would probe the cosmological birefringence
or constrain the coupling strength, $\beta_{F\tF}$. Another proposed
method is the CMB polarization coming from the last scattering
surface~\cite{Lue,LLN,LLN14,LN17} and hence upper limits on the rotated angles
derived from CMB polarization data have been reported~\cite{planckCPV}.
In the next section, we will study CMB polarization in the presence
of the cosmological birefringence induced by DE and DM, taking into account the
birefringence perturbation.

\section{Radiative transfer with rotating polarization plane}
\label{QUsec}

Thomson scatterings of anisotropic radiation by free electrons
give rise to the linear polarization, which is usually described
by the Stokes parameters
$Q(\eta, \vec{x})$ and $U(\eta, \vec{x})$~\cite{chandrasekar1960}.
When the polarization plane is rotated by an angle $\alpha$,
the Stokes parameters are transformed into
\be
\left ( \begin{array}{c} Q'\\U' \end{array} \right)=
\left( \begin{array}{cc}
\cos2\alpha&\sin2\alpha\\-\sin2\alpha&\cos2\alpha
\end{array} \right)
\left ( \begin{array}{c} Q\\U \end{array} \right).
\ee
Hence, the angular velocity~(\ref{omegaetax}) due to the cosmological birefringence
of the dark sector would lead to the temporal rate of change of the Stokes parameters:
\be
\dot Q\pm i \dot U = \mp i2\omega \left( Q\pm i U \right),
\label{QUeq}
\ee
where the dot denotes $d/d\eta$. In standard cosmology, the time evolution of the
polarization perturbation is governed by the Boltzmann equation~\cite{BEa}.
When there is a physical mechanism which rotates the polarization plane,
the evolution equations for the Fourier modes
of the Stokes parameters are modified to
\ba
&&\dot{\Delta}_{Q\pm iU}({\vec k},\eta) + ik\mu \Delta_{Q\pm iU}
({\vec k},\eta)=
n_e\sigma_T a(\eta) \Biggl[ -\Delta_{Q\pm iU}({\vec k},\eta) +
  \nonumber \\
&& \left .\sum_{m}\sqrt{\frac{6\pi}{5}}
\ _{\pm 2} Y_2^m ({\hat n}) S_{P}^{(m)}({\vec k},\eta) \right]
\mp i 2 \fr{1}{\sqrt{(2\pi)^{3}}}
\int d{\vec k}'\,\tilde{\omega}({\vec k}-{\vec k}',\eta) \Delta_{Q\pm iU}({\vec k}',\eta),
\label{fpsnew}
\ea
where the Fourier transform is
\be
\omega(\eta,\vec{x}) =
\fr{1}{\sqrt{(2\pi)^{3}}} \int \tilde{\omega}(\vec{k} , \eta)
e^{i \vec{k} \cdot \vec{x}} d^3 {\vec k} \, ,
\ee
$\mu={\hat n}\cdot{\hat k}$ is the cosine of the
angle between the CMB photon direction and the Fourier wave
vector, $n_e$ is the number density of free electrons, and $\sigma_T$
is the Thomson cross section.
$\ _sY_l^m$ are spherical harmonics with spin-weight $s$, where
$m=0,\pm 1, \pm 2$ correspond, respectively, to scalar, vector, and
tensor perturbations with the axis of $\ _sY_l^m$ aligned with the
wave vector ${\vec k}$. $S_P^{(m)}$ is the source term for
generating polarization, being composed of the quadrupole
components of the temperature and polarization perturbations
$S_P^{(m)}({\vec k},\eta) \equiv \Delta_{T,2}^{(m)}({\vec k},\eta)+
12\sqrt{6}\Delta_{+,2}^{(m)}({\vec k},\eta)+12\sqrt{6}\Delta_{-,2}^
{(m)}({\vec k},\eta)$. Here we have followed the notation in
Ref.~\cite{Liu}. We have expanded the temperature ($\Delta_T$)
and polarization ($\Delta_{Q\pm iU}$) perturbations in terms of
$Y_l^m$ and $\ _{\pm2} Y_l^m$~\cite{NP}, respectively,
and denoted the expansion coefficients by
$\Delta_{T,{l}}^{(m)}$ and $\Delta_{\pm,{l}}^{(m)}$.

To consider the effect of the
perturbation of the scalar field on the cosmological birefringence,
we can decompose the scalar field as
\be \phi(\eta,\vec{x}) =
\bar{\phi}(\eta) + \delta \phi (\eta, \vec{x}) \label{phi} \, ,
\ee
where $\bar{\phi}$ is the vev and $\delta \phi$
is the perturbed part of the scalar field. For the metric perturbation,
we adopt the synchronous gauge:
\be
ds^2= a^2(\eta) \left\{d\eta^2- \left[\delta_{ij}+h_{ij}(\eta, {\vec x})\right]dx^i dx^j\right\}
\label{gmunu}\, .
\ee
From the field equation~(\ref{phieq}), we obtain the mean field evolution as
\be
\ddot{\bar\phi} + 2 {\cal H} \dot{\bar\phi} +
\fr{a^2}{M^2} \fr{\partial V}{\partial \bar\phi}
- \fr{1}{a^2 M^2} \fr{\partial B_{F \tilde{F}}}{\partial \bar\phi}
(\vec{E} \cdot \vec{B}) = 0\, ,\label{phi0eq}
\ee
where ${\cal H} \equiv \dot{a} / a$. The perturbation equation is given by
\be
\frac{\partial^2\delta\phi}{\partial\eta^2}
+ 2 {\cal H}  \frac{\partial\delta\phi}{\partial\eta}
- \vec{\nabla}^2 \delta \phi +
\fr{a^2}{M^2} \fr{\partial^2 V}{\partial \bar\phi^2} \delta \phi
- \fr{1}{a^2 M^2} \fr{\partial^2 B_{F \tilde{F}}}{\partial
\bar\phi^2} (\vec{E} \cdot \vec{B}) \delta \phi =
-{1\over2} \frac{\partial h}{\partial\eta} \frac{d\bar\phi}{d\eta}\, ,
\label{deltaphieq}
\ee
where $h$ is the trace of $h_{ij}$.
Doing the Fourier expansion of the perturbed scalar field and the trace,
\ba
\delta \phi(\eta,\vec{x}) &=&
\fr{1}{\sqrt{(2\pi)^{3}}} \int \delta \phi_{\vec k}(\eta) e^{i
\vec{k} \cdot \vec{x}} d^3 {\vec k} \, , \label{deltaphi1} \\
h(\eta,\vec{x}) &=&
\fr{1}{\sqrt{(2\pi)^{3}}} \int h_{\vec k}(\eta) e^{i
\vec{k} \cdot \vec{x}} d^3 {\vec k} \, , \label{trace}
\ea
and using Eq.~(\ref{omegaetax}), we obtain $\omega$ and its Fourier transform
$\tilde{\omega}$ in Eq.~(\ref{fpsnew})
to the first order in perturbation as
\ba
\omega(\eta,\vec{x})&=&-{1\over2}\fr{\partial^2B_{F\tF}}{\partial {\bar\phi}^2}
\dot{\bar\phi}\;{\delta\phi}
-{1\over2}\fr{\partial B_{F\tF}}{\partial {\bar\phi}}
\left(\fr{\partial {\delta\phi}}{\partial\eta}+
\vec{\nabla} {\delta\phi} \cdot \hat{n}\right),\label{omegapert}\\
\tilde{\omega}({\vec k}, \eta)&=& -{1\over2}\fr{\partial^2B_{F\tF}}{\partial {\bar\phi}^2}
\dot{\bar\phi}\;{\delta\phi}_{\vec k}
-{1\over2}\fr{\partial B_{F\tF}}{\partial {\bar\phi}}
\left(\dot{\delta\phi}_{\vec k}+
i{\vec k}\cdot \hat{n}\, \delta\phi_{\vec k}\right).
\label{tildeomega}
\ea
Neglecting the back-reaction in the last term of Eq.~(\ref{deltaphieq}),
we have the equation of motion for $\delta\phi_{\vec k}$:
\be
\ddot{\delta\phi}_{\vec k}
+ 2 {\cal H} \dot{\delta\phi}_{\vec k} + \left( k^2+
\fr{a^2}{M^2} \fr{\partial^2 V}{\partial \bar\phi^2}\right) {\delta\phi}_{\vec k}=
-{1\over2} {\dot h_{\vec k}} {\dot{\bar\phi}}\, .
\label{fourierphi}
\ee
Eqs.~(\ref{fpsnew}), (\ref{tildeomega}), and (\ref{fourierphi}) form a full set of
equations for the time evolution of CMB polarization in the presence of inhomogeneous
birefringence. Then we can solve this set of equations by a numerical method
in conjunction with publicly available CMB numerical codes such as CMBFAST and CAMB~\cite{SZ}. 

\section{Free-streaming approximation}
\label{freestream}

However, we may make use of the fact that the primary CMB polarization is generated on the last scattering surface
at the time of decoupling and on the rescattering surface in the epoch of reionization.
Under the assumption that the CMB polarization is mostly
generated before the birefringence-induced rotation of the polarization plane
takes place,  we can neglect the Thomson scattering term
in Eq.~(\ref{fpsnew}) and only consider the rotation of the polarized CMB coming from
the last scattering surface or from the rescattering surface. This enables us to
derive useful analytic solutions for checking with the numerical results.

In the absence of Thomson scatterings, the propagation of the CMB photons
is simply the free streaming with the plane of polarization
being rotated due to the birefringence. From Eq.~(\ref{QUeq}), we find
that the linear polarization at the present time is given by the
line-of-sight integral,
\ba
(Q\pm U)({\hat n})&=&(Q\pm U)({\hat n}, \eta_s)\,e^{\mp i2\alpha({\hat n})}\, , \label{qualpha}\\
\alpha({\hat n})&=&\int_{\eta_s}^{\eta_0}\omega[\eta,(\eta_0-\eta){\hat n}]d\eta \, ,
\label{alphalos}
\ea
where we have written $\vec x=r{\hat
n}=(\eta_0-\eta){\hat n}$ and $\eta_s$ denotes the time when the
primary CMB polarization is generated on the last scattering
surface or the rescattering surface. Using Eq.(\ref{omegapert}),
the line-of-sight integral simply gives
\be
\alpha({\hat n})=\left.{1\over2}\fr{\partial B_{F\tF}}{\partial {\bar\phi}}(\eta)\;
\delta\phi[\eta,(\eta_0-\eta){\hat n}]\right\vert_{\eta_0}^{\eta_s}.
\label{alphaetas}
\ee
Hence, the vev part gives a homogeneous rotated angle of the polarization plane,
\be
{\bar\alpha}={1\over2}\fr{\partial B_{F\tF}}{\partial {\bar\phi}}(\eta)\left[\bar{\phi}(\eta_s)-\bar{\phi}(\eta_0)\right].
\label{baralpha}
\ee
As such, the rotation-induced CMB power spectra are given by
\ba
&&C_l^B=C_l^E \sin^2 2{\bar\alpha}, \quad C_l^{TB}=C_l^{TE} \sin 2{\bar\alpha},\nonumber\\
&&C_l^{EB}={1\over2}C_l^{E} \sin 4{\bar\alpha}.
\label{clbaralpha}
\ea
For the inhomogeneous perturbation, we perform the spherical
harmonics expansion: 
\ba
(Q\pm U)({\hat n})&=& \sum_{lm} a_{\pm,l}^m \ _{\pm2} Y_l^m ({\hat n})\, ,\\
(Q\pm U)({\hat n}, \eta_s)&=& \sum_{lm} a_{\pm,l}^m (\eta_s) \ _{\pm2} Y_l^m ({\hat n})\, ,\\
\alpha({\hat n})&=& \sum_{lm} \alpha_l^m Y_l^m ({\hat n})\, .
\ea
Since it is expected that $\alpha({\hat n}) \ll 1$, we expand the phase term in Eq.~(\ref{qualpha}).
Then, keeping only the lowest order, we find that
\be
a_{\pm,l}^m = a_{\pm,l}^m (\eta_s) \mp i2 \sum_{l_1 m_1 l_2 m_2}
a_{\pm,l_1}^{m_1} (\eta_s) \alpha_{l_2}^{m_2} \int d{\hat n}
\ _{\pm2} Y_l^{m*} ({\hat n}) \ _{\pm2} Y_{l_1}^{m_1} ({\hat n}) Y_{l_2}^{m_2} ({\hat n})\, .
\label{almalpha}
\ee
Thus, the polarization power spectra are given by the primary polarization
power spectra convolved with the rotation power spectrum $\alpha_l^m $.
To evaluate $\alpha_l^m $, we use the Fourier expansion in Eq.~(\ref{deltaphi1})
and do the expansion,
\be
e^{i\vec{k} \cdot \vec{x}} = 4\pi\sum_{lm}\, i^l j_l(kr) Y_l^{m*}(\hat k) Y_l^m (\hat n)\, .
\ee
Then, we obtain from Eq.~(\ref{alphaetas}) that
\be
\alpha_l^m= i^l{1\over\sqrt{2\pi}} \fr{\partial B_{F\tF}}{\partial {\bar\phi}}(\eta_s)
\int d^3{\vec k}\, Y_{l}^{m*}(\hat k)\,
{\delta\phi}_{\vec k}(\eta_s) j_l[k(\eta_0-\eta_s)]\, .
\label{alphalm}
\ee
The remaining integral in Eq.~(\ref{almalpha}) can be expressed
in terms of the Wigner $3$-$j$ symbols through the general formula~\cite{varsh},
\ba
&&\int d{\hat n}
\ _{s} Y_{l}^{m} ({\hat n}) \ _{s_1} Y_{l_1}^{m_1} ({\hat n}) \ _{s_2} Y_{l_2}^{m_2} ({\hat n})
\nonumber\\
&=& \sqrt{\frac{(2l+1)(2l_1+1)(2l_2+1)}{4\pi}}
\left( \begin{array}{ccc} l &l_1 &l_2\\ -s &-s_1 &-s_2\end{array} \right)
\left( \begin{array}{ccc} l &l_1 &l_2\\  m &m_1 &m_2\end{array} \right)\, .
\label{3j}
\ea
Isotropy in the mean guarantees the ensemble averages:
\begin{eqnarray}
\left<a_{T,l'}^{m'*} a_{T,l}^{m}\right>&=&C_l^T\delta_{l'l}\delta_{m'm}\, ,
\nonumber \\
\left<a_{\pm,l'}^{m'*} a_{\pm,l}^{m}\right>&=&(C_l^E+C_l^B)\delta_{l'l}
\delta_{m'm}\, , \nonumber \\
\left<a_{+,l'}^{m'*} a_{-,l}^{m}\right>&=&(C_l^E-C_l^B)
\delta_{l'l}\delta_{m'm}\, , \nonumber \\
\left<a_{T,l'}^{m'*} a_{\pm,l}^{m}\right>&=&-C_l^{TE}\delta_{l'l}\delta_{m'm}\, ,
\nonumber \\
\left<\alpha_{l'}^{m'*} \alpha_{l}^{m}\right>&=&C_l^{\alpha}\delta_{l'l}\delta_{m'm}\, ,
\label{enav}
\end{eqnarray}
and both $TB$ and $EB$ mode power spectra, unlike the homogeneous case, vanish.
Hence, Eq.~(\ref{almalpha}) implies that the rotation-induced $B$-mode power spectrum is given by
\be
C_l^B={1\over\pi}\sum_{l_1,l_2} (2l_1+1)(2l_2+1) C_{l_1}^E(\eta_s) C_{l_2}^{\alpha}
\left( \begin{array}{ccc} l &l_1 &l_2\\  2 &-2 & 0\end{array} \right)^2\, ,
\ee
where we have assumed a negligible primary $B$ mode and used the relation,
\be
(2l+1)\sum_{m_1,m_2}
\left( \begin{array}{ccc} l' &l_1 &l_2\\  m' &m_1 &m_2\end{array} \right)
\left( \begin{array}{ccc} l &l_1 &l_2\\  m &m_1 &m_2\end{array} \right)
= \delta_{l'l}\delta_{m'm}\, .
\label{sum3j}
\ee

Before we find the rotation power spectrum $C_l^{\alpha}$,
it is interesting to re-consider Eq.~(\ref{alphalos}) in terms of the
Fourier mode ${\tilde\omega}(\vec k, \eta)$.
Here we simply assume $B_{F\tF} = \beta_{F\tF} \phi$. As such, we have
\be
\alpha_l^m= \sqrt{2\over\pi} \sum_{l_1 m_1} \int d^3{\vec k}\, d{\hat n}\int_{\eta_s}^{\eta_0}d\eta\,
{\tilde\omega}(\vec k, \eta)\,i^{l_1} j_{l_1}[k(\eta_0-\eta)]
Y_{l_1}^{m_1*}(\hat k) Y_{l_1}^{m_1}(\hat n) Y_{l}^{m*}(\hat n)\, ,
\label{alphalmlong}
\ee
where Eq.~(\ref{tildeomega}) implies that
\be
\tilde{\omega}({\vec k}, \eta)=-{1\over2} \beta_{F\tF}
\left(\dot{\delta\phi}_{\vec k}+
i{\vec k}\cdot \hat{n}\, \delta\phi_{\vec k} \right).
\ee
In Eq.~(\ref{alphalmlong}), $\alpha_l^m$ depends on
${\tilde\omega}(\vec k,\eta)$ which contains two different terms.
Let us split $\alpha_l^m$ into
$\alpha_l^m=\alpha_{1l}^m+\alpha_{2l}^m$, where
the first term is the temporal contribution and
the second term is the gradient. It is then straightforward
to show that the temporal term is given by
\be
\alpha_{1l}^m= i^l{1\over\sqrt{2\pi}} \beta_{F\tilde{F}}
\int d^3{\vec k}\, Y_{l}^{m*}(\hat k) \left\{
{\delta\phi}_{\vec k}(\eta_s) j_l[k(\eta_0-\eta_s)]
-k \int_{\eta_s}^{\eta_0}d\eta\,
{\delta\phi}_{\vec k}(\eta) j_{l}'[k(\eta_0-\eta)]\right\}\, .
\label{alphalm1}
\ee
For the gradient term $\alpha_{2l}^m$, we first expand
\be
\delta\phi_{\vec k}=\sum_{lm}\delta\phi_{kl}^m Y_{l}^{m}(\hat k)\, ,
\quad
{\hat k}\cdot{\hat n}=\frac{4\pi}{3} \sum_{m=-1}^{1} Y_1^m(\hat k) Y_1^{m*}(\hat n)\, .
\ee
This enables us to do the integration over the solid angles:
\ba
&&\sum_{m_1}\int d{\hat k}\, d{\hat n}\, ({\hat k}\cdot{\hat n})
Y_{l_2}^{m_2}(\hat k) Y_{l_1}^{m_1*}(\hat k) Y_{l_1}^{m_1}(\hat n) Y_{l}^{m*}(\hat n)
\nonumber\\
&=& (-1)^{(1+l+l_1)}(2l_1+1)
\left( \begin{array}{ccc} 1 &l &l_1\\  0 & 0 & 0\end{array} \right)^2
\delta_{l_2 l}\delta_{m_2 m}\, ,
\ea
where we have used the Wigner $3$-$j$ symbols~(\ref{3j}), the relation in Eq.~(\ref{sum3j}),
and the properties of the symbols given by
\be
\left( \begin{array}{ccc} l &l_1 &l_2\\  m &m_1 &m_2\end{array} \right)
=(-1)^{l+l_1+l_2}\left( \begin{array}{ccc} l &l_2 &l_1\\  m &m_2 &m_1\end{array} \right)
=(-1)^{l+l_1+l_2}\left( \begin{array}{ccc} l &l_1 &l_2\\  -m &-m_1 &-m_2\end{array} \right)\, .
\ee
Hence we obtain that
\ba
\alpha_{2l}^m &=& {1\over\sqrt{2\pi}}\beta_{F\tilde{F}}
\int d^3{\vec k}\, Y_{l}^{m*}(\hat k)
\, k \int_{\eta_s}^{\eta_0}d\eta\,
{\delta\phi}_{\vec k}(\eta)\, \times\nonumber\\
&&i (-1)^l\sum_{l_1} (-i)^{l_1} j_{l_1}[k(\eta_0-\eta)] (2l_1+1)
\left( \begin{array}{ccc} 1 &l &l_1\\  0 & 0 & 0\end{array} \right)^2\, ,
\label{alphalm2}
\ea
where the summation over $l_1$ satisfies the triangular inequalities,
$l-1\le l_1 \le l+1$ (i.e. $l_1=l-1,l,l+1$).

\subsection{Initial conditions of quintessence perturbation}

To proceed the calculation, we need to specify the initial conditions for the perturbation
${\delta\phi}_{\vec k}(\eta)$ at an early time $\eta_i$. Let us consider the energy density
and pressure of the quintessence,
\be
\rho_{\phi}=-\frac{M^2}{2} g^{\mu\nu} \partial_\mu \phi \partial_\nu \phi + V(\phi)\, ,\quad
p_{\phi}=-\frac{M^2}{2} g^{\mu\nu} \partial_\mu \phi \partial_\nu \phi - V(\phi)\, .
\ee
From Eqs.~(\ref{phi}) and (\ref{gmunu}), their means and fluctuations are then given by
\ba
&&\bar\rho_{\phi}=\frac{M^2}{2a^2} \dot{\bar\phi}^2 + V(\bar\phi)\, ,\quad
\bar p_{\phi} =\frac{M^2}{2a^2} \dot{\bar\phi}^2 - V(\bar\phi)\, ,\\
&&\delta\rho_{\phi}(\vec k,\eta)=\frac{M^2}{a^2} \dot{\bar\phi}\,\dot{\delta\phi_{\vec k}} +
\frac{\partial V}{\partial{\bar\phi}} \delta\phi_{\vec k}\, ,\quad
\delta p_{\phi}(\vec k,\eta)=\frac{M^2}{a^2} \dot{\bar\phi}\,\dot{\delta\phi_{\vec k}} -
\frac{\partial V}{\partial{\bar\phi}} \delta\phi_{\vec k}\, .
\ea
Inflation creates a nearly scale-invariant primordial power spectrum
of adiabatic density perturbations in all light fields. It means that
the entropy perturbation for the entire fluid, just after inflation, vanishes:
\be
T \delta s = \delta p - \frac{\dot{\bar p}}{\dot{\bar\rho}} \delta \rho = 0\, .
\label{TdS}
\ee
This enables us to find the relation between $\dot{\delta\phi}$ and $\delta\phi$.
From the fact that long-wavelength fluctuation modes are frozen outside the horizon,
we also set $\dot{\delta\phi}(\eta_i)= 0$.

In Ref.~\cite{dave}, the authors considered two initial conditions, the so-called smooth
and adiabatic cases. The former is that $\dot{\delta\phi}=\delta\phi= 0$; the latter is
that $\delta\rho_\phi/{\bar\rho_\phi}=\delta\rho_r/{\bar\rho_r}=(4/3)\delta\rho_m/{\bar\rho_m}$,
where the last two are the radiation and matter energy densities respectively.
The adiabatic case has this form because the EOS of quintessence equals to that of radiation
($w_{\phi} = w_{r} = 1/3$) in the early radiation-dominated epoch.
It was shown that the CMB anisotropy power spectrum is insensitive to the choice of initial conditions.
The difference in the anisotropy power spectrum is much smaller than the cosmic variance limit in both cases.
We can use the initial condition, $\delta\rho_\phi/{\bar\rho_\phi} \simeq 10^{-16}$.
This implies that $\delta\phi(\eta_i)\simeq 10^{-16}$ and $\dot{\delta\phi}(\eta_i)
= 0$. It was also shown that the isocurvature initial condition can be ignored
because the isocurvature perturbation decays with time~\cite{Lee}.
Thus, we can just consider the adiabatic initial condition for our calculation.

Given a quintessence potential, we solve for the time evolution of the mean field $\bar\phi$ in
Eq.~(\ref{phi0eq}) and the perturbation ${\delta\phi}_{\vec k}$
in Eq.~(\ref{fourierphi}). Then, using Eq.~(\ref{alphalm}), we obtain the rotation power spectrum as
\be
C_l^{\alpha}=\left<|\alpha_l^m|^2\right>=\frac{1}{2\pi}
\left[\fr{\partial B_{F\tF}}{\partial {\bar\phi}}(\eta_s)\right]^2
\int {dk}{k^2} \left\{ {\delta\phi}_k (\eta_s)\, j_l[k(\eta_0-\eta_s)]\right\}^2\,,
\ee
where we have assumed that
\be
\left<\delta\phi_{\vec k}(\eta_s)\,\delta\phi_{\vec k'}(\eta_s)\right> =
 {\delta\phi}_k^2(\eta_s)\, \delta({\vec k}-{\vec k}')\, .
\ee

\subsection{Nearly a cosmological constant}

When the source term in the right-hand side of the equation of motion~(\ref{fourierphi})
can be neglected, the perturbation ${\delta\phi}_{\vec k}$ can be factored into
\be
{\delta\phi}_{\vec k}(\eta)={\delta\phi}_{\vec k,i}\,f_k(\eta)\, ,
\ee
where ${\delta\phi}_{\vec k,i}$ is the initial perturbation amplitude
and $f_k(\eta)$ satisfies
\be
\ddot{f_k}
+ 2 {\cal H} \dot{f_k} + \left( k^2+
\fr{a^2}{M^2} \fr{\partial^2 V}{\partial \bar\phi^2}\right) f_k=0\, ,
\label{fourierfk}
\ee
with $f_k(\eta_i)=1$.
If the quintessence is nearly massless or its effective mass is less than
the present Hubble parameter, the dark energy would behave
just like a cosmological constant and the perturbation
${\delta\phi}_{\vec k}$ would be dispersive. In this case,
$f_k(\eta)$ is a dispersion factor which can be cast into $f_k(\eta)=f(k\eta)$.
For a super-horizon mode with $k\eta\ll 1$, $f(k\eta)=1$; the factor then
oscillates with a decaying envelope once the mode enter the horizon.
Let us define the initial power spectrum $P_\phi(k)$
of the quintessence perturbation by
\be
\left<\delta\phi_{\vec k,i}\delta\phi_{\vec k',i}\right> = \frac{2\pi^2}{k^3}
P_\phi(k) \, \delta({\vec k}-{\vec k}')\, .
\ee
Then, using Eq.~(\ref{alphalm}), we find that the rotation power spectrum is given by
\be
C_l^{\alpha}=\pi\left[\fr{\partial B_{F\tF}}{\partial {\bar\phi}}(\eta_s)\right]^2
\int \frac{dk}{k} \, P_\phi(k)
\left\{ f_k(\eta_s) j_l[k(\eta_0-\eta_s)]\right\}^2\,.
\ee
When separating the temporal and gradient contributions, given respectively in
Eq.~(\ref{alphalm1}) and Eq.~(\ref{alphalm2}), we have
\be
C_l^{\alpha}=\left<|\alpha_l^m|^2\right>=\left<|\alpha_{1l}^m|^2\right>+
\left<|\alpha_{2l}^m|^2\right>+2\,{\rm Re}\left<\alpha_{1l}^{m*}\alpha_{2l}^m\right>\, ,
\ee
where
\ba
\left<|\alpha_{1l}^m|^2\right>&=&\pi\beta_{F\tilde{F}}^2
\int \frac{dk}{k} \, P_\phi(k) \times \nonumber \\
&&\left\{ f_k(\eta_s) j_l[k(\eta_0-\eta_s)]
-k \int_{\eta_s}^{\eta_0}d\eta\,f_k(\eta) j_{l}'[k(\eta_0-\eta)]\right\}^2\, ,\\
\left<|\alpha_{2l}^m|^2\right>&=&\pi\beta_{F\tilde{F}}^2
\int \frac{dk}{k} \, P_\phi(k) \times \nonumber \\
&&\left| k \int_{\eta_s}^{\eta_0}d\eta\,
f_k(\eta) \sum_{l_1} (-i)^{l_1} j_{l_1}[k(\eta_0-\eta)] (2l_1+1)
\left( \begin{array}{ccc} 1 &l &l_1\\  0 & 0 & 0\end{array} \right)^2\right|^2\, ,\\
2\,{\rm Re}\left<\alpha_{1l}^{m*}\alpha_{2l}^m\right>&=&2\pi\beta_{F\tilde{F}}^2
\int \frac{dk}{k} \, P_\phi(k) \times \nonumber \\
&&\left\{ f_k(\eta_s) j_l[k(\eta_0-\eta_s)]
-k \int_{\eta_s}^{\eta_0}d\eta\,f_k(\eta) j_{l}'[k(\eta_0-\eta)]\right\}\times \nonumber \\
&&\left\{ k \int_{\eta_s}^{\eta_0}d\eta\,
f_k(\eta) \left[ (2l+3)j_{l+1}[k(\eta_0-\eta)]
\left( \begin{array}{ccc} 1 &l &l+1\\  0 & 0 & 0\end{array} \right)^2\right.\right.- \nonumber \\
&&\left.\left. (2l-1)j_{l-1}[k(\eta_0-\eta)]
\left( \begin{array}{ccc} 1 &l &l-1\\  0 & 0 & 0\end{array}\right)^2\right]\right\}\, .
\ea

\subsection{Dark matter perturbation}

The time evolution of $\phi$ is determined by its scalar potential, $V(\phi)={1\over2} m^2M^2\phi^2$.
When $\phi$ begins to oscillate at $a=a_{\rm osc}$, the energy density of the scalar condensate is given by
\begin{equation}
\rho=m^2 M^2\bar{\phi}^2=m^2 M^2\bar{\phi}_i^2 \left(\frac{a_{\rm osc}}{a}\right)^3,
\end{equation}
where $\bar{\phi}_i$ is the initial vev. Hence, the perturbation is
\begin{equation}
{\delta\phi}={1\over2} \bar{\phi} \frac{\delta\rho}{\rho}
=\frac{\sqrt{3}}{2}\Omega_{\rm DM}^{1\over2} \frac{H_0}{m} (1+z)^{3\over2} \frac{\delta\rho}{\rho},
\end{equation}
where $\Omega_{\rm DM}$ and $H_0$ each take the present values when $a=a_0=1$.
Here $\delta\equiv {\delta\rho}/{\rho}$ is assumed to be the adiabatic DM density perturbation.
In terms of their perturbation power spectra, we have
\begin{equation}
\Delta_{\delta\phi}^2(k,\eta)
={3\over 4}\Omega_{\rm DM} \left(\frac{H_0}{m}\right)^2 (1+z)^3 \Delta_\delta^2(k,\eta) ,
\end{equation}
Hence the rotation power spectrum is given by
\be
C_l^{\alpha}= \pi \left[\fr{\partial B_{F\tF}}{\partial {\bar\phi}}(\eta_s)\right]^2
\int \frac{dk}{k} \Delta_{\delta\phi}^2(k,\eta_s) \, j_l[k(\eta_0-\eta_s)]^2\,.
\label{clalpha}
\ee

\section{Rotation power spectra and CMB anisotropy-polarization}
\label{powerspec}

In this section, we summarize the main results obtained in Refs.~\cite{LLN,LLN14,LN17}.
The rotation power spectra due to DE and DM birefringence are generally deviated from scale-invariance. 
The rotation power spectrum for the scalar DE is red-tilted while that for the scalar DM is blue.
As such, for the scalar DE the birefringence induced $B$-mode polarization power spectra follow the shape of the $E$-mode
power spectrum, while the induced $B$-mode power spectra for the scalar DM are blue.

\begin{figure}[htbp]
\centerline{\psfig{file=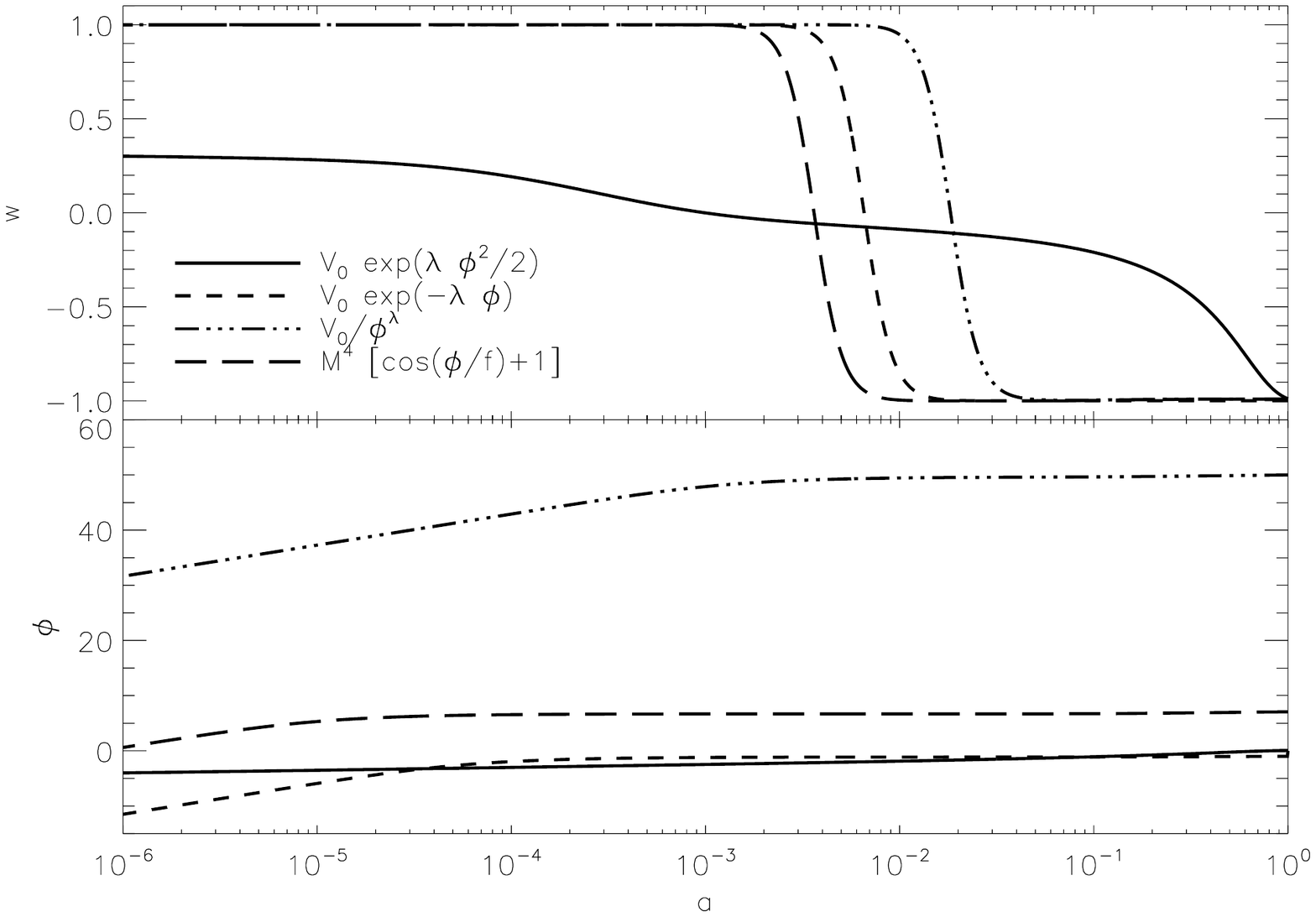, width=14cm}
}\caption{Evolution of the equation of state, $w_{\phi}$, and the vev $\bar{\phi}$ in scalar dark energy or quintessence models,
whose respective parameters are chosen as to obtain the evolution consistent with current observational data.}
\label{fig1}
\end{figure}

\begin{figure}[htbp]
\centerline{\psfig{file=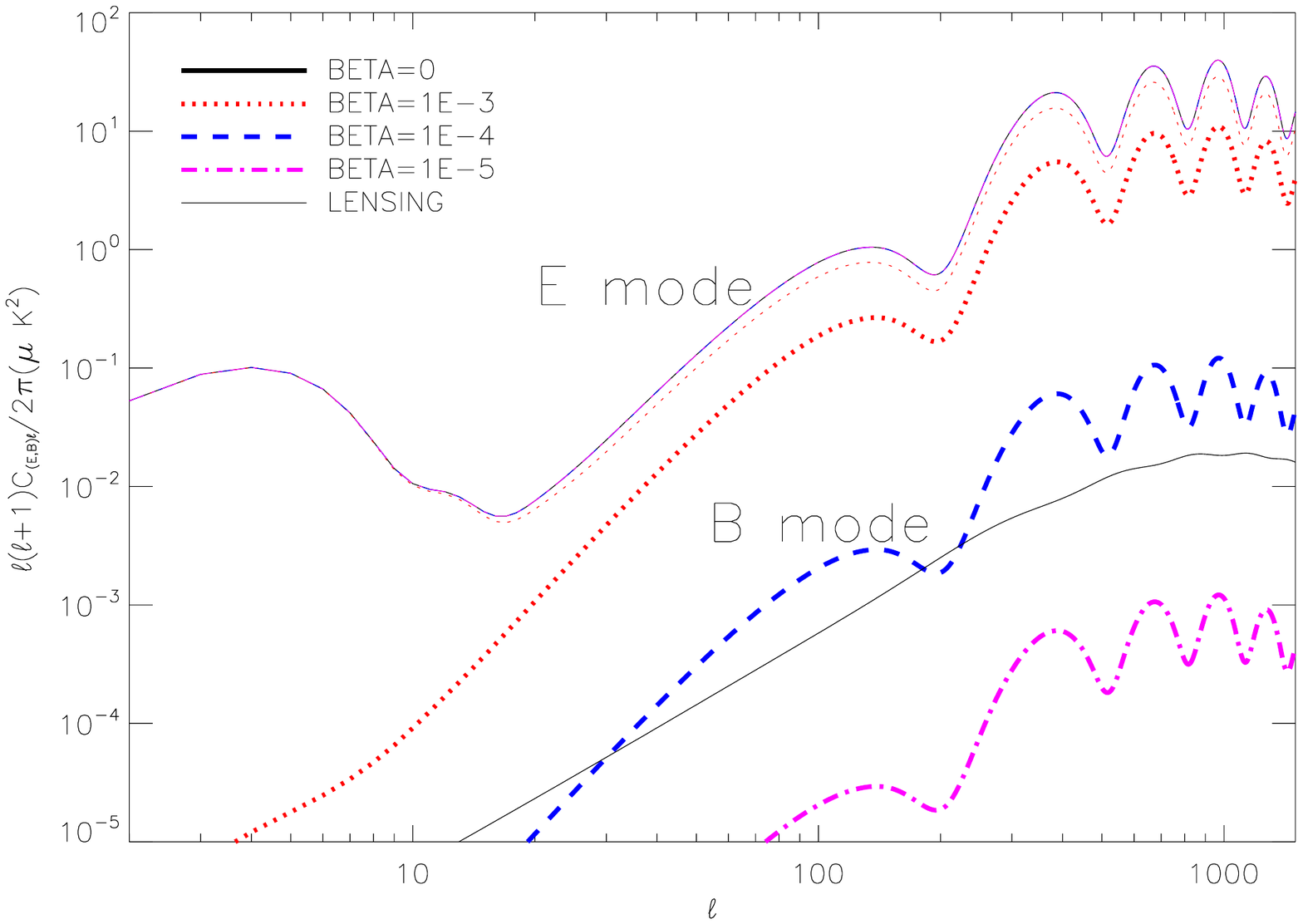, width=8cm}
\psfig{file=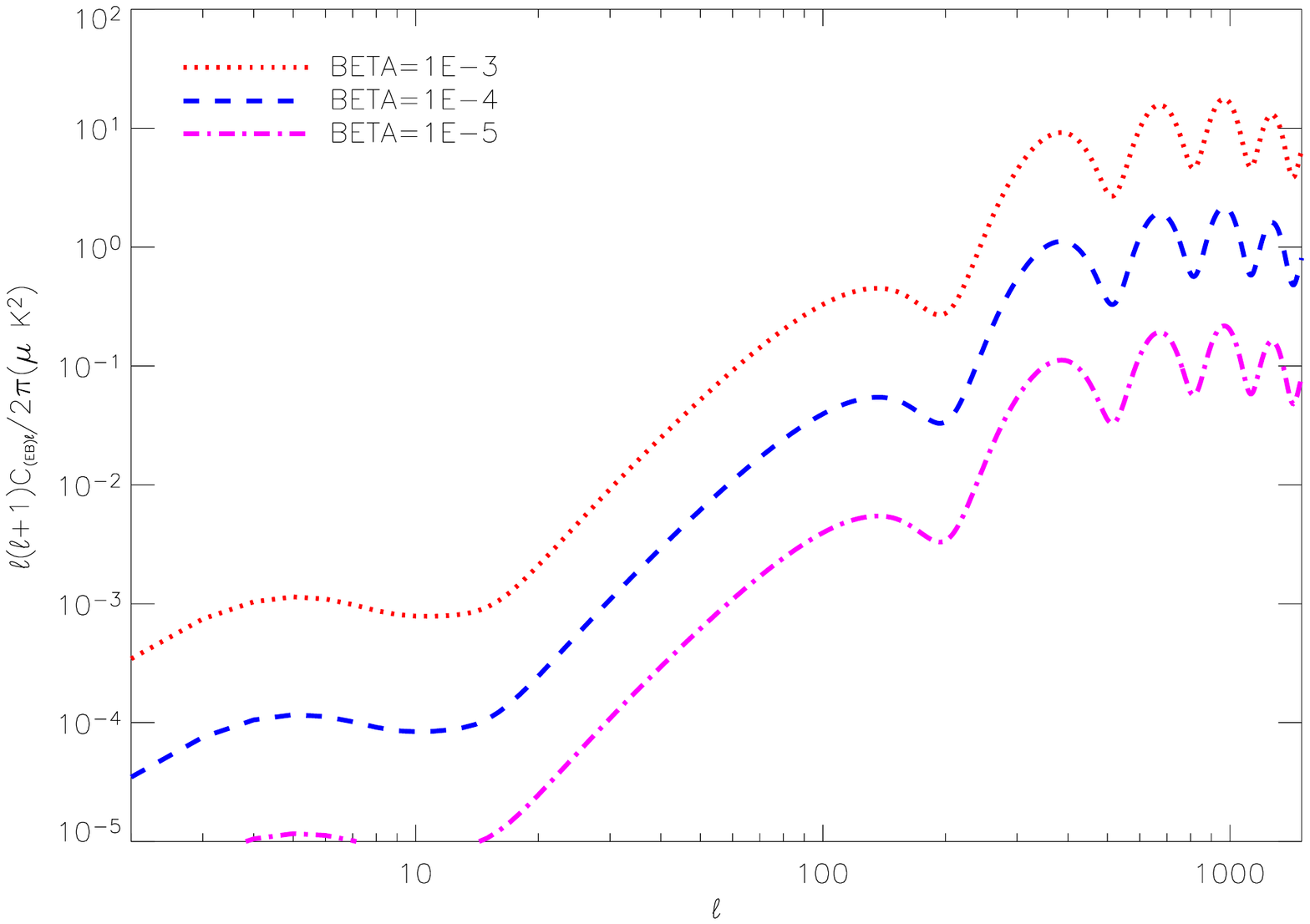, width=8cm}} 
\caption{$E$, $B$ (left panel; the lower three thick curves are $B$ modes) and
$EB$ (right panel) mode power spectra from the cosmological quintessence
birefringence with different coupling strength.}
\label{fig2}
\end{figure}

\begin{figure}[htbp]
\centerline{\psfig{file=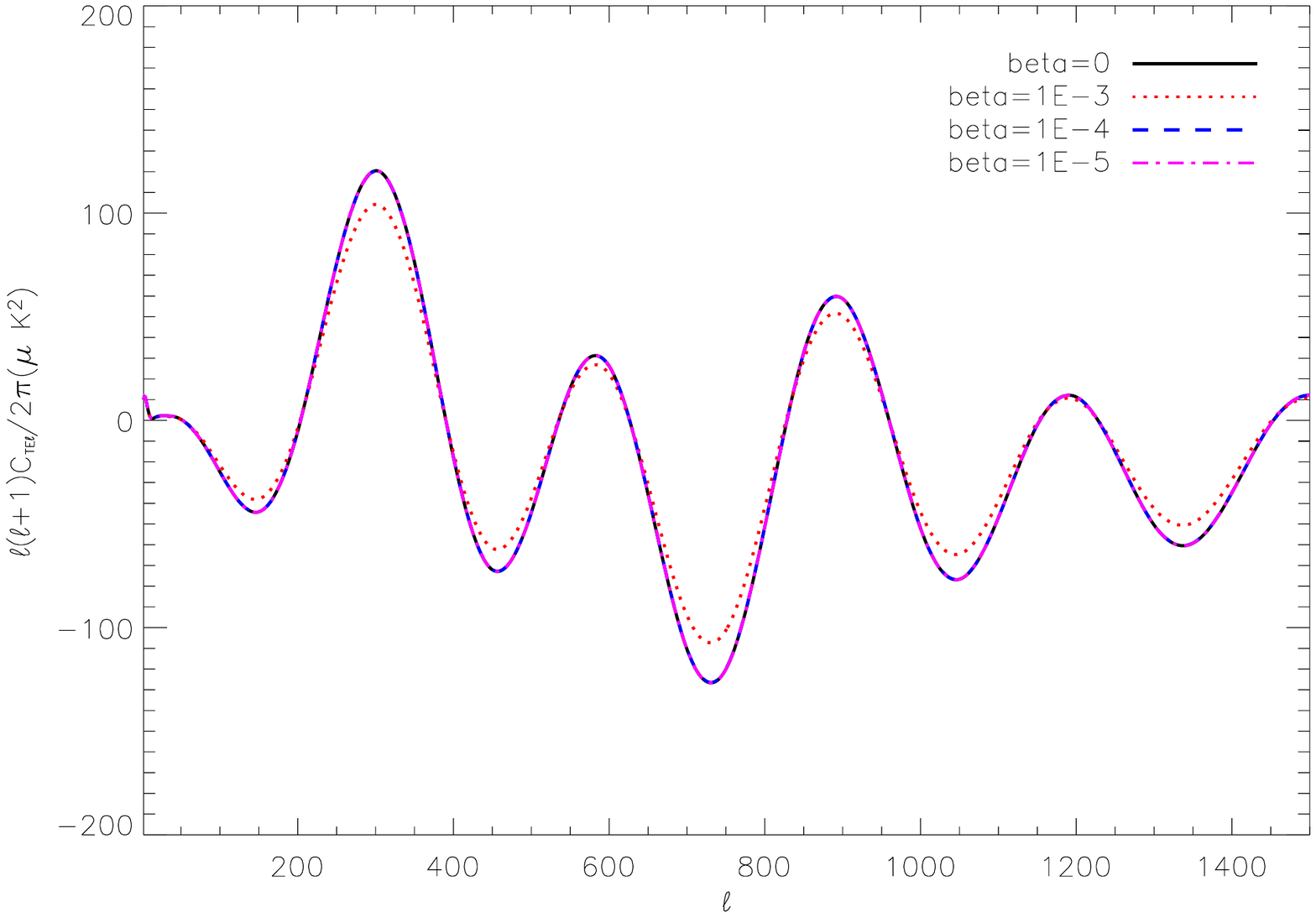, width=8cm}
\psfig{file=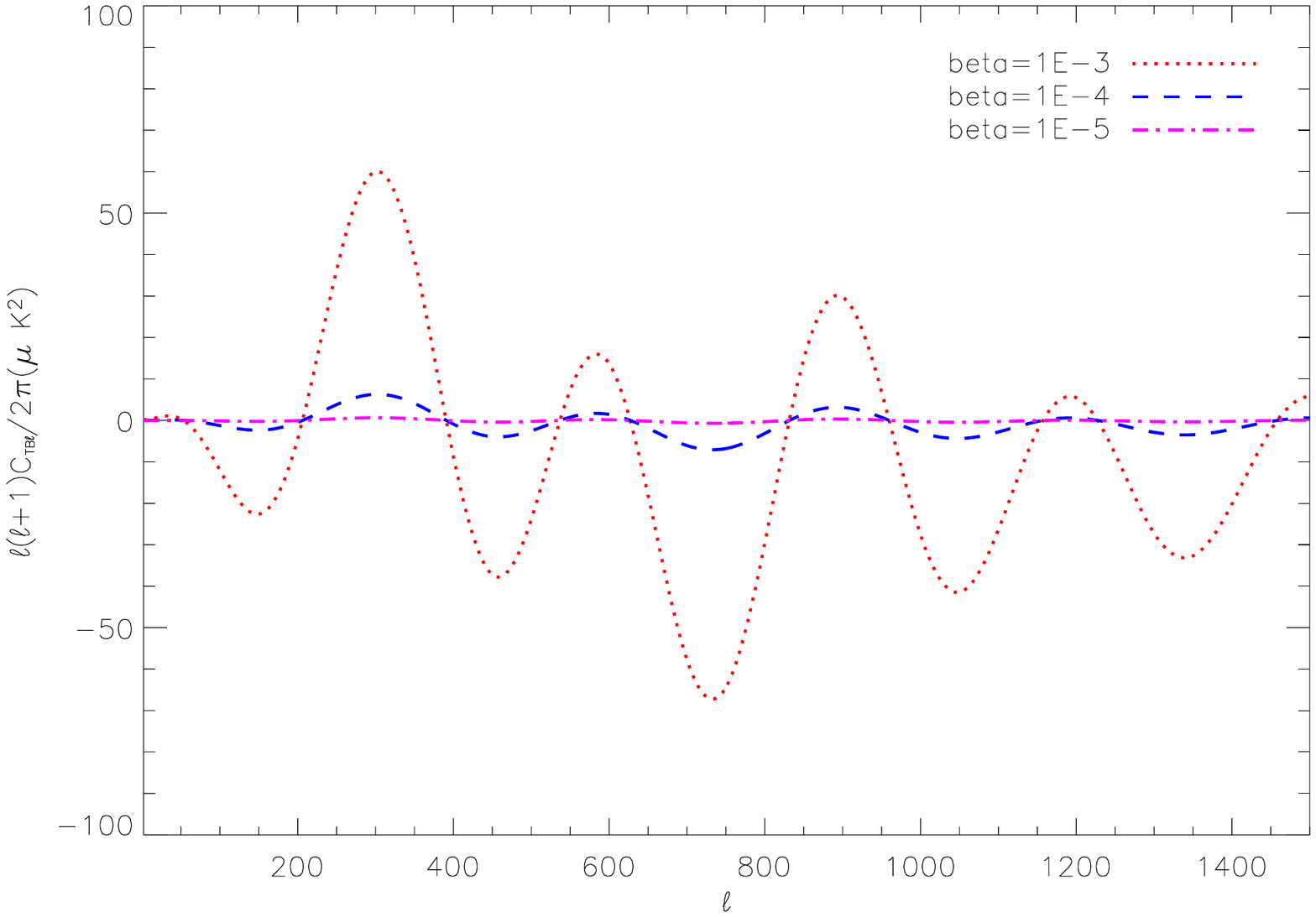, width=8cm}}
\caption{$TE$
(left panel) and $TB$ (right panel) mode power spectra from the
cosmological quintessence birefringence with different coupling strength.}
\label{fig3}
\end{figure}

In Fig.~\ref{fig1}, we plot the evolution of $w_{\phi}$ and $\bar{\phi}$ in some representative
scalar dark energy models, adjusting each model parameters such that the evolution is consistent
with current observational data. The obtained results are quite similar to
one another. Here we just provide the results of a typical exponential potential,
$V(\phi)=V_0 \rm{exp}(\lambda \phi^2/2)$, with $\lambda=5$.
The left panel of Fig.~\ref{fig2} shows the $E$ and $B$ mode power spectra with the coupling
strength $\beta_{F\tilde{F}}$ ranging from $10^{-5}$ to $10^{-3}$.
The $EB$ mode power spectrum is shown in the right panel. 
The shapes of the $B$ and $EB$ mode power spectra
basically follow the standard $E$ mode except the reionization bump
on large scales. The $E$ mode power on small scales mainly comes from the recombination
epoch at $z \sim 1100$. On the other hand, the boosting power on
large scales comes from reionization epoch when the CMB photons are
rescattered by free electrons at $z \sim 10$. From
Eq.~(\ref{baralpha}) and the evolution of $\bar{\phi}$, we find that the
integrated rotation angle from the reionization epoch is much
smaller than that from the recombination epoch. Therefore, there is
much less power converted from $E$ mode to $B$ mode on large scales
than small scales. We also show the power spectrum of the
lensing-induced $B$ mode in Fig.~1 by a thin solid curve for
comparison. Fig.~\ref{fig3} shows the $TE$ and $TB$ power spectra for different coupling
strength. The induced $B$-mode power spectrum
due to the perturbation $\delta\phi$
for both adiabatic and smooth initial conditions is tiny, with a peak about $10^{-31} \mu K^2$
for $\beta_{F\tilde{F}}=1.0$, and insensitive to initial conditions. Considering the combined $TB$ and $EB$ data, 
the {\em Planck} team in Ref.~\cite{planckCPV} found that the rotation angle 
$\bar{\alpha}=0.31\pm0.05({\rm stat.})\pm0.28({\rm syst.})$, which is compatible with no parity violation and 
is dominated by the systematic uncertainty in the orientation of Planck's polarization-sensitive bolometers.
This constraint implies that $\beta_{F\tilde{F}} < 10^{-3}$. 

\begin{figure}[htbp]
\centerline{\psfig{file=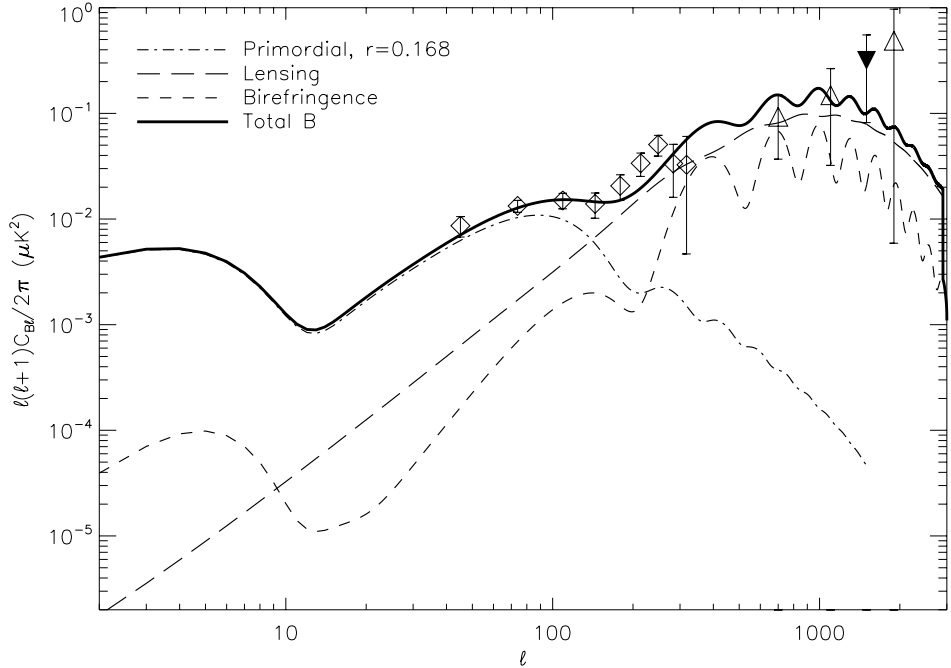, width=14cm}
} \caption{
Cosmological birefringence induced $B$-mode power spectrum through the
perturbed nearly massless scalar dark energy with $A\beta_{F\tilde{F}}^2=0.0046$ 
(short-dashed).
Also shown are the theoretical power spectra of lensing induced
$B$ modes (long-dashed) and gravity-wave induced $B$ modes (dot-dashed) with $r=0.168$. The thick solid curve is the best-fitting
averaged $B$-mode band powers that are the sum of these three $B$-mode power spectra convolved with the BICEP2 ($l<400$) and the POLARBEAR ($l>400$) window functions. BICEP2 data~\cite{bicep2} (diamonds) and POLARBEAR data~\cite{polarbearB} (triangles and an inverted solid triangle representing the absolute value of a negative band) are shown.}
\label{fig4}
\end{figure}

\begin{figure}[htbp]
\centerline{\psfig{file=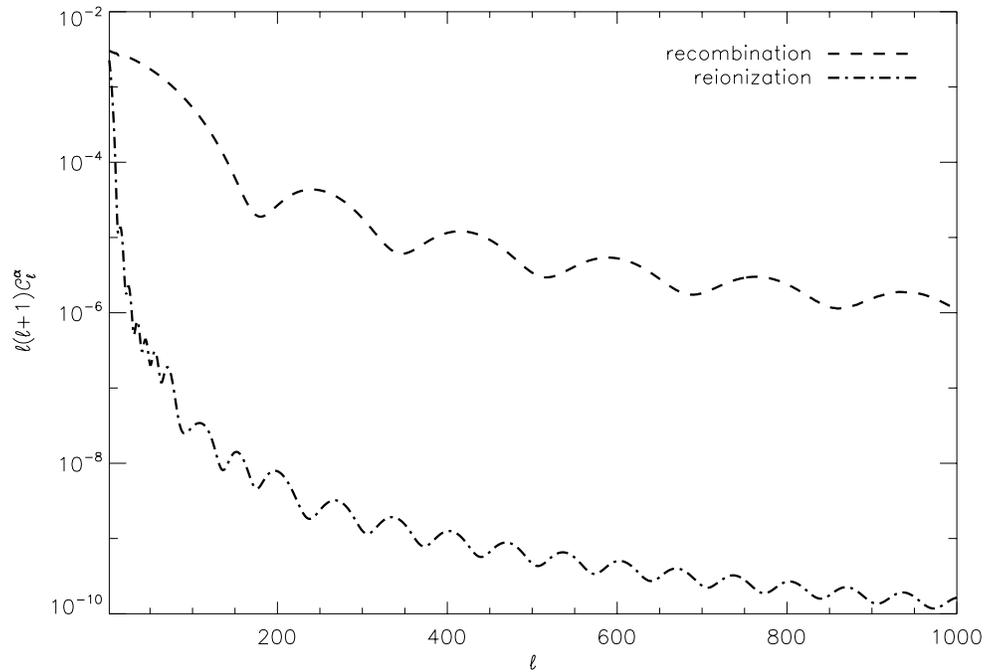, width=14cm}
} \caption{Rotation power spectra at the recombination and the reionization for nearly massless scalar dark energy with $A\beta_{F\tilde{F}}^2=0.0046$.}
\label{fig5}
\end{figure}

For the case that $\phi$ is nearly massless, we take the initial power spectrum as
$P_{\delta\phi} (k)=Ak^{n-1}$, where $A$ is a constant amplitude squared and $n$ is the spectral index.
Using a scale-invariant spectrum ($n=1$) and a combined constant parameter $A\beta_{F\tilde{F}}^2$,
we have tuned the values of the tensor-to-scalar ratio $r$ and $A\beta_{F\tilde{F}}^2$, by fixing the other cosmological parameters to the 
best-fit values of the {\em Planck} 6-parameter LCDM model~\cite{planck16}, to best fit BICEP2 and POLARBEAR data as shown in Fig.~\ref{fig4}. 
We have also produced the rotation power spectra for the recombination and
the reionization with $A\beta_{F\tilde{F}}^2=0.0046$ in Fig.~\ref{fig5}.

\begin{figure}[t]
\centerline{\psfig{file=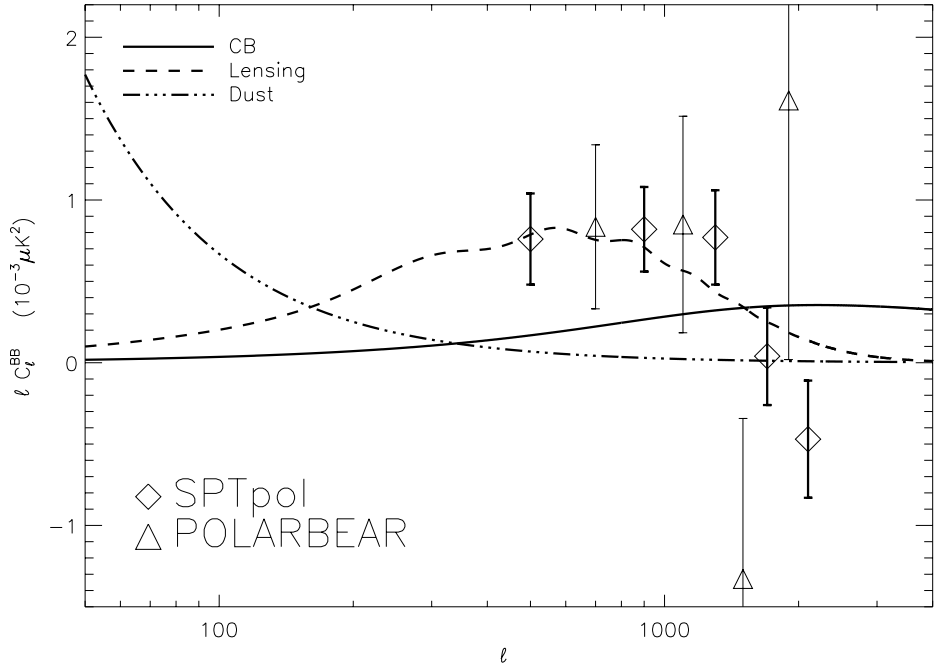, width=14cm}} 
\caption{Scalar dark matter induced $B$-mode power spectrum with $A_{\rm CB}=8\times 10^{15}$ (solid).
Also shown are the power spectra of lensing induced
$B$ modes with $A_L=1.07$ (dashed) and dust $B$ modes (dot-dashed). Overlaid are POLARBEAR data~\cite{polarbearB} (triangles) and SPTpol data~\cite{sptB} (diamonds).}
\label{fig6}
\end{figure}

\begin{figure}[htbp]
\centerline{\psfig{file=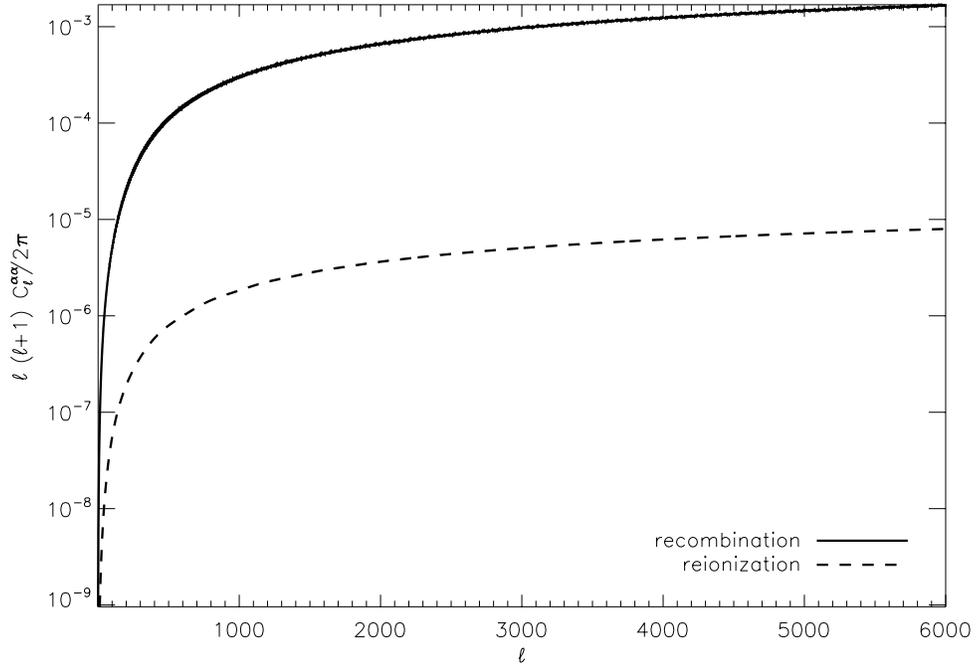, width=14cm}}
\caption{Rotation power spectra at the recombination and the reionization for scalar dark matter with $A_{\rm CB}=8\times 10^{15}$.}
\label{fig7}
\end{figure}

For the scalar DM case, we define the cosmic birefringence parameter $A_{\rm CB}\equiv\beta_{F\tilde{F}}^2 (10^{-22}{\rm eV}/m)^2$. 
Figure~\ref{fig6} shows the lensing and birefringence $B$-mode power spectra with the lensing parameter $A_L=1.07$ (where $A_L=1$ for the LCDM model) and $A_{\rm CB}=8\times 10^{15}$ respectively. The birefringence $B$ modes dominate the polarization power for $l>1400$; therefore, measurements of $B$-mode polarization at sub-degree scales are critical for probing cosmic birefringence induced by scalar dark matter. Figure~\ref{fig7} shows the blue-tilted rotation power spectra for the recombination and the reionization with $A_{\rm CB}=8\times 10^{15}$.

\section{Conclusions}
\label{conclus}

If dark energy is birefringent, there would be a wide window for us to see its
properties through measurements of the CMB polarization.
Dynamical dark energy would rotate $E$-polarization into $B$-polarization, thus leaving
cosmic parity-violating $TB$ and $EB$ correlation as well as rotation-induced $B$-modes.
These $B$-mode power spectra are similar to the lensing $B$-mode and the gravity-wave induced $B$-mode,
so the parity violation is crucial to distinguishing between them. Even though dark energy is
indeed a cosmological constant, its perturbation can still generate a rotation-induced $B$-mode
power spectrum while conserving the cosmic parity. In this case, it is a big challenge to do
the separation of different $B$-mode signals. It is apparent that the rotation-induced
$B$-mode has acoustic oscillations but to detect them will require next-generation experiments.

Furthermore, there can be a new source of CMB $B$-mode polarization induced by birefringence fluctuations of ultralight axion-like dark matter.
The power spectrum of this birefringence $B$-mode polarization peaks at sub-degree angular scales and may be at a level detectable in on-going CMB lensing $B$-mode searches such as ACTpol, POLARBEAR, and SPTpol experiments. Interestingly, it may dominate over the lensing $B$-mode power spectrum at higher-$l$ range. Thus, it would be very important to make precise measurements of $B$-mode polarization at sub-degree scales to disentangle the two $B$-mode signals. 
The current experimental sensitivity in measuring $l C_l^{BB}$ is of order $10^{-3} {\rm\mu K^2}$, which is at the same level of the $B$-mode signals. 
In future CMB-S4 polarization experiments, the sensitivity will be tremendously improved to $\sim 10^{-6} {\rm\mu K^2}$ for $l<5000$~\cite{cmbs4}, so consistency of sub-degree $B$ modes with the lensing of $E$ modes will test the present model in a well-defined way. 

In principle, one may use de-lensing methods~\cite{delensing}
or lensing contributions to CMB bi-spectra~\cite{bispectra} to single out the lensing $B$ mode.
In addition, de-rotation techniques can be used to remove the rotation-induced
$B$ modes~\cite{kamion09}. More investigations along this line should be done in order to disentangle all possible $B$-mode sources,
importantly before confirming the detection of the genuine primordial $B$-mode polarization.

\section*{Acknowledgments}

This work was supported in part by the Ministry of Science and Technology, Taiwan, ROC under the Grants No. MOST104-2112-M-001-039-MY3 (K.W.N.) and No. MOST105-2112-M-032 -002 (G.C.L.).



\end{document}